# Observation of two distinct superconducting domes under pressure in tetragonal FeS

J. Zhang[1,*], F. L. Liu[1,2,*], T. P. Ying[1,*], N. N. Li[2], Y. Xu[1], L. P. He[1], X. C. Hong[1], Y. J. Yu[1], M. X. Wang[1], J. Shen[1,3], W. G. Yang[2,4,†], and S. Y. Li[1,3,‡]

[1] *State Key Laboratory of Surface Physics, Department of Physics, and Laboratory of Advanced Materials, Fudan University, Shanghai 200433, China*

[2] *Center for High Pressure Science and Technology Advanced Research (HPSTAR), Shanghai 201203, China*

[3] *Collaborative Innovation Center of Advanced Microstructures, Nanjing 210093, China*

[4] *High Pressure Synergetic Consortium (HPSynC), Geophysical Laboratory, Carnegie Institution of Washington, Argonne, IL 60439, USA*

[*] *These authors contribute equally to this work.*

**As the simplest iron-based superconductor, FeSe forms a tetragonal structure with transition temperature $T_c \approx 8$ K. With assistance of pressure, or other techniques, $T_c$ can be greatly enhanced, even to above liquid nitrogen temperature. The newly discovered superconducting tetragonal FeS ($T_c \approx 4.5$ K), a sulfide counterpart of FeSe, promotes us on its high pressure investigation. The transport and structure evolution of FeS with pressure have been studied. A rapid suppression of $T_c$ and vanishing of superconductivity at 4.0 GPa are observed, followed by a second superconducting dome with a 30% enhancement in maximum $T_c$. An onsite tetragonal to hexagonal phase transition occurs around 7.0 GPa, followed by a broad pressure range of phase coexistence. The residual deformed tetragonal phase is considered as the source of second superconducting dome. The observation of two superconducting domes in iron-based superconductors poses great challenges for understanding their pairing mechanism.**

As the second family of high-$T_c$ superconductor, the iron-based superconductors (IBSs) have been extensively studied in recent years [1-3]. All the reported IBSs share a common $Fe_2X_2$ (X = As, P or Se)-layered structure unit, which possesses an anti-PbO-type (anti-litharge type) atom arrangement. The $Fe_2X_2$ layers consist of edge-shared $FeX_4$ structure, where X ions form a distorted tetrahedral arrangement around the Fe ions. Both theoretical and experimental works suggest that the detailed crystal structure, including bond angle and anion height, plays a key role in the superconductivity of IBSs [4-6].

Among those IBSs, the tetragonal FeSe with the simplest structure has attracted much attention recently [7]. Though the $T_c$ of bulk FeSe is modestly low ($\approx$ 8 K) [7], it can be enhanced through various methods. By applying high pressure, the $T_c$ of bulk FeSe can reach up to 36.7 K [8,9]. Through intercalation, surface K dosing, or ionic liquid gating, one can also enhance the $T_c$ to above 40 K [10-14]. Surprisingly, the $T_c$ can be further enhanced above 60 K by growing monolayer FeSe thin film on $SrTiO_3$ substrate [15-19], and even above 100 K on Nb-doped $SrTiO_3$ substrate [20]. It was believed that the enhanced electron-phonon coupling at interface is crucial in the further enhancement of $T_c$ [21,22]. The angle-resolved photoemission spectroscopy (ARPES) studies reveal two distinct electronic band structures for these FeSe-based superconductors. The Fermi surface of the undoped bulk FeSe consists of hole pockets around Γ and electron pockets around M [18,23-26]. However, for intercalated $(Li_{0.8}Fe_{0.2})OHFeSe$, monolayer FeSe thin film, and surface K dosed FeSe single crystal or film, the Fermi surface consists of only electron pockets, which apparently results from electron doping [16-19,27-30]. For undoped bulk FeSe, the superconducting pairing symmetry is most likely $s_\pm$ type with sign reversal between the hole and electron pockets [31], while for monolayer FeSe/$SrTiO_3$, plain $s$-wave superconductivity was suggested by scanning tunneling microscopy (STM) study [32]. Due to the large variety of FeSe-based superconductors with a wide range of $T_c$, clarifying the superconducting mechanism will be a major step towards solving the issue of high temperature superconductivity in IBSs.

Very recently, superconductivity with $T_c \approx$ 4.5 K at ambient pressure was reported for FeS [33], the sister compound of FeSe. As an important Earth Science material, FeS has been extensively studied including high-pressure works [34], but superconductivity had not been observed until high-quality stoichiometric tetragonal

FeS was successfully synthesized by low-temperature hydrothermal method [33]. The tetragonal FeS has the same crystal structure as the tetragonal FeSe, and their electronic structures are also quite similar based on first principle calculation [35]. A slight difference between them is that FeSe undergoes a phase transition to orthorhombic structure at 90 K [36], while FeS remains its tetragonal structure down to 10 K [37]. Interestingly, recent thermal conductivity and specific heat measurements suggested nodal superconductivity in FeS [38,39].

Here we present *in situ* high-pressure electrical transport and synchrotron x-ray diffraction measurements on tetragonal FeS single crystals. Upon applying pressure, two distinct superconducting domes are observed. The first dome manifests a continuous decrease of $T_c$ with increasing pressure, ending around 4.0 GPa. Then a second superconducting dome emerges from 5.0 GPa and lasts to 22.3 GPa, with an over 30% increasing in $T_c$ ($\approx$ 6.0 K) from the highest $T_c$ in the first dome, and the $T_c$ stays almost constant in the entire second dome. Comparing to the two dome superconducting behavior of $(K/Tl/Rb)_xFe_{2-y}Se_2$ reported earlier [40], the superconducting pressure range is much wider, meaning the superconducting phase is much more sustainable with pressure. For the crystal structure under pressure, a hexagonal phase starts to set in at around 7.0 GPa, and there is a large coexisting pressure range of tetragonal and hexagonal phases. Based on the drop in $R(T)$ curves and the structure refinement results of mixture phase region, we believe that the second superconducting dome is originated from the residual deformed tetragonal phase of FeS.

**Results**

**Characterization at ambient pressure.** We first demonstrate that our FeS samples synthesized by de-intercalation of K from $KFe_{2-x}S_2$ precursor are single crystals. The inset of Fig. 1a is a photo of as-grown FeS crystals. Figure 1a shows a typical x-ray diffraction (XRD) pattern, in which only the (00*l*) Bragg peaks show up, indicating the crystals are well oriented along the *c* axis. Further XRD measurement on these crystals shows sharp single crystal diffraction spots (Fig. 1b). Flat and grain-boundary-free surface was observed with Scanning Electron Microscopy (inset of Fig. 1b). Therefore, the single crystalline nature of our FeS samples is confirmed.

Figure 1c shows a typical low-temperature dc magnetization of FeS single crystals. The superconducting transition is at about 4.1 K, and there is no positive ferromagnetic background in the normal state. The temperature dependence of resistivity $\rho(T)$ at ambient pressure is plotted in Fig. 1d. The absence of a resistivity anomaly in the normal state suggests no structural phase transition, which is different from FeSe single crystal [36]. The low-temperature resistivity between 5 and 50 K can be well described by the Fermi liquid theory, $\rho(T) = \rho_0 + AT^2$, giving $\rho_0 = 6.07$ μΩ cm and $A = 2.0 \times 10^{-3}$ μΩ cm/K$^2$. The residual resistivity ratio, RRR = $\rho(298\text{ K})/\rho_0 = 40$, is much larger than that reported previously for FeS flakes [33]. The inset of Fig. 1d displays an enlarged view around the superconducting transition, from which $T_c^{onset} \approx 4.7$ K and $T_c^{zero} \approx 4.3$ K are obtained. $T_c^{onset}$ is determined as the temperature where the resistivity deviates from the normal-state behavior, while $T_c^{zero}$ as the temperature where the resistivity drops to zero. In the following discussions, we use $T_c^{onset}$ as $T_c$.

**$T_c$ evolution under pressure.** The temperature dependence of resistance up to 4.0 GPa is plotted in Fig. 2a, where the resistance is normalized to its value at 15 K at each pressure. Initially, the $T_c$ is suppressed rapidly with increasing pressure, consistent with previous reports (below 2.2 GPa) by other groups [41,42]. Here, we observe $T_c$ eventually disappears at 4.0 GPa. The superconductivity in this region is so sensitive to pressure that the transition broadens and the resistance does not drop to zero even under 0.86 GPa. Figure 2b and 2c show the normalized resistance curves above 5.0 GPa. The drop of resistance re-emerges below 4.5 K at 5.0 GPa, implying the arise of another superconducting phase. This resistance drop exists in a wide pressure range from 5.0 to 22.3 GPa, and maximum $T_c$ reaches 6 K, a 30% enhancement from the highest value in the first dome. With further increasing pressure, the resistance drop vanishes and the $R(T)$ curve exhibits a semiconducting behavior.

To make sure the resistance drop under high pressure represents a superconducting transition, we applied magnetic field to the low-temperature resistance measurements at 19.0 GPa. As shown in Fig. 2d, the resistance drop is gradually suppressed to lower temperature with increasing field, which demonstrates that it is indeed a superconducting transition. The inset of Fig. 2d plots the reduced temperature $T/T_c$ dependence of the upper critical field $H_{c2}$. The data can be fitted to the generalized Ginzburg-Landau model: $H_{c2}(T) = H_{c2}(0)(1 - t^2)/(1 + t^2)$, where $t =$

$T/T_c$. According to the fit, $H_{c2}(0) \approx 0.81$ T is obtained.

**Crystal structure evolution under pressure.** *In situ* high-pressure synchrotron powder XRD measurements were utilized to study the structural evolution of FeS with pressure. Figure 3a displays the obtained XRD patterns under various pressures at room temperature. At the lowest pressure (1.0 GPa), the pattern can be well characterized as the tetragonal phase. From 7.2 to 9.2 GPa, a set of new peaks emerges with increasing intensity, while the intensity of the original peaks decreases. This indicates a structural transition and the coexistence of two different phases. The peaks from the low-pressure phase cannot be distinguished above 10.1 GPa, and the high-pressure phase remains stable up to 38.1 GPa.

Based on the Rietveld refinements, the low-pressure structure can be well indexed in the tetragonal space group *P4/nmm*, with the lattice parameters $a = 3.650$ Å and $c = 4.940$ Å at 1.0 GPa. Comparing to the ambient pressure FeS structure [33], the values of $a$ and $c$ decrease slightly due to the shrinkage of lattices under pressure. On the high-pressure side, the hexagonal space group *P-62c* is found to be the optimal structure when we refine the XRD data above 10.1 GPa. The corresponding two crystal structures are shown in Fig. 3b and 3c for *P4/nmm* and *P-62c*, respectively. Similar pressure-induced structural transition from a tetragonal to hexagonal phase was also observed in FeSe, with a wide pressure range for two-phase coexistence [8]. The pressure dependence of the lattice parameters $a$, $c$, and unit cell volume is plotted in Fig. 3d, 3e, and 3f, respectively. These lattice parameters show an abrupt change when the tetragonal structure transforms to the hexagonal one. The unit cell volume of the hexagonal phase is 13% (7.2 GPa) smaller than that of the tetragonal phase, which reveals the increase of the sample density, as expected.

**Temperature-pressure phase diagram.** We summarize our experimental results in Fig. 4. Figure 4a shows the pressure dependence of phase content around the structural transition. The hexagonal phase first appears around 7.2 GPa and its content increases rapidly with pressure. The original tetragonal phase occupies a small portion (~ 3%) at 9.2 GPa and is hardly distinguishable through the refinement above 10.1 GPa. The temperature-pressure (*T-P*) phase diagram is summarized in Fig. 4b. The superconductivity is rapidly suppressed by pressure in the first superconducting dome (SC-I), while it re-emerges in a wide pressure range from 5.0 to 22.3 GPa, manifesting as a second superconducting dome (SC-II) with a maximum $T_c$ of 6.0 K around 16.1 GPa. Upon further compression, FeS remains the hexagonal structure and behaves as a semiconductor.

Since the two-phase (tetragonal + hexagonal) coexisting region (7.2 - 9.2 GPa) lies inside the second superconducting dome (5.0 - 22.3 GPa), we try to identify which phase is responsible for SC-II. Firstly, well inside the second dome, e.g. $P$ = 13.0 and 16.1 GPa, the sample is dominated by the hexagonal phase, but the resistance drop is only a few percent. This suggests that the SC-II should not come from the major hexagonal phase. Secondly, all the IBSs have manifested superconductivity in either tetragonal or orthorhombic phase so far [1-3]. Thirdly, hexagonal FeS shows semiconducting resistance behavior under ambient and high pressure [43,44]. Therefore, it is very likely that the SC-II arises from the remaining tetragonal phase of FeS, which coexists with the hexagonal phase up to about 22.3 GPa. For the XRD patterns with only a few percent of tetragonal phase, it is beyond the refinement capability to distinguish it from the major hexagonal phase, thus the Rietveld refinements beyond 10.3 GPa have ignored the contribution of tetragonal phase.

**Discussions**

The two domes we observe here in FeS are quite different from the single dome observed in the $T$-$P$ phase diagram of $A$Fe$_2$As$_2$ ($A$ = alkaline-earth metals) and $R$FeAsO ($R$ = rare-earth metals) [45]. For the sister compound FeSe, its $T$-$P$ phase diagram has only one superconducting dome with the maximum $T_c$ = 36.7 K at 8.9 GPa [8]. Since sulfur atom has a smaller radius than selenium atom, FeS can be considered as FeSe under chemical pressure and is easier to compress. In this sense, the rapid $T_c$ suppression in the first dome of FeS below 4 GPa may correspond to the high-pressure side of the superconducting dome observed in FeSe. So far, there is no report on the second superconducting dome in the $T$-$P$ phase diagram of FeSe. We will discuss the possible connection with the structure study by XRD below.

Two superconducting domes in the $T$-$P$ phase diagram were reported previously in two other IBS systems [39, 46-50]. For (K/Tl/Rb)$_x$Fe$_{2-y}$Se$_2$, the $T_c$ has a maximum value of 32 K at 1 GPa within the first dome, and a maximum $T_c$ of 48.7 K in the second dome between 9.8 and 13.2 GPa [46]. The reason for this re-emergency of superconductivity under high pressure is still unknown. For $KFe_2As_2$, the $T_c$ exhibits a V-shaped dependence under $P < 3$ GPa, which was suggested as an indication of pairing symmetry change [46-48]. Similar behavior was also observed in $RbFe_2As_2$ and $CsFe_2As_2$ [49]. Upon further compressing $KFe_2As_2$, a structural transition takes place from the tetragonal to collapsed tetragonal phase around 16 GPa, and meanwhile the carrier characteristic is changed [50]. The $T_c$ is greatly enhanced to 12 K in the collapsed tetragonal phase, which may be caused by strong electronic correlations in this phase [50].

We notice that previous studies suggested a close relationship between the structure of $Fe_2X_2$ layer and $T_c$ in IBSs [4,5,51,52]. For the FeAs-based "1111" superconductors, $T_c$ becomes maximum when the $FeAs_4$ structure forms a regular tetrahedron with As–Fe–As bond angle = 109.47° [4,5]. For $FeSe_xTe_{1-x}$ ($x = 0.5 \sim 0.875$), the $T_c$ increases with increasing $X$-Fe-$X$ ($X$ = Se or Te) bond angle $\alpha$ [5]. For FeSe under pressure, the $T_c$ and the unit-cell metrics $(a + b)/2c$ show the same evolution trend with pressure [51]. While for $FeSe_{0.8}S_{0.2}$ under pressure, the $T_c$ anti-correlates with the anion height $h$, as it decreases with increasing $h$ and increases with decreasing $h$ [52].

Based on the Rietveld refinements of the XRD data, we perform a similar analysis on the detailed crystal structure of FeS under high pressure. Figure 5a shows the structure of $Fe_2S_2$ layer, where S-Fe-S bond angles $\alpha, \beta$ and Fe-S-Fe bond angles $\gamma, \delta$, as well as anion height are marked. The pressure dependence of anion height, obtained from $z$ position of the sulfur atom and the lattice parameter $c$, is shown in Fig. 5b. The anion height basically decreases with increasing pressure, with a small anomaly around 3.0 GPa. The Fe-S bond length decreases monotonically with the applied pressure, as plotted in Fig. 5c. Figure 5d and 5e show the pressure dependence of the four bond angles $\alpha$ and $\beta$, $\gamma$ and $\delta$. A V-shaped pressure dependence of $\alpha$ can be clearly observed, where $\alpha$ decreases with pressure below 3.0 GPa, but recovers back above 4.5 GPa. Coincidentally, the $T_c$ shares a very similar evolution trend with $\alpha$ in this pressure range, as shown in Fig. 5f. This indicates that the bond angle $\alpha$ may play an important role in determining the $T_c$ of FeS. Besides $\alpha$ angle, the other angles $\beta, \gamma$ and $\delta$ have similar trends as they are connected by the structural symmetry. Despite that there is a close correlation between $T_c$ and crystal structure for FeS and above mentioned IBSs, more experimental and theoretical studies are desired to reveal the underlying physics of these superconducting domes.

In summary, we demonstrate two distinct superconducting domes in the temperature-pressure phase diagram of the newly discovered superconductor FeS by means of high-pressure resistance measurements. The *in situ* high-pressure XRD results reveal a phase transition from pristine tetragonal to a hexagonal structure in a broad pressure range. The superconductivity in both domes originates from tetragonal FeS phase. We point out a close correlation between the $T_c$ and the S-Fe-S bond angle $\alpha$ in the tetragonal structure with pressure. The observation of two superconducting domes in FeS, together with similar results reported earlier in other IBSs, poses great challenges for understanding the pairing mechanism of IBSs.

## Methods

**Sample synthesis and characterizations.** FeS single crystals were synthesized by de-intercalation of K from $KFe_{2-x}S_2$ precursor by hydrothermal method [53]. First, $KFe_{2-x}S_2$ single crystals were grown by self-flux method [54]. Then several pieces of $KFe_{2-x}S_2$ single crystals, Fe powder, $CN_2H_4S$, and NaOH in mole ratio 1:1:2:20 were loaded into a Teflon-lined stainless steel autoclave together with 15 ml deionized water. The autoclave was then sealed and kept at 130℃ for 24 hours. Large single crystals could be harvested by rinsing the products in deionized water several times and drying under vacuum. X-ray diffraction (XRD) was carried out at room temperature using Bruker D8 Advance diffractometer with Cu $K_\alpha$ radiation ($\lambda$ = 1.5408 Å). Scanning electron Microscopy (SEM) images were taken on an Electron Probe Microanalyzer (Shimadzu, EPMA-1720). Single crystal XRD of FeS was carried out on a Bruker SMART Apex (II) diffractometer (Mo $K_\alpha$ radiation, $\lambda$ = 0.71073 Å). The dc magnetization was measured in a Superconducting Quantum Interference Device (SQUID, Quantum Design). Electrical resistivity measurement at ambient pressure was performed in $^4$He and $^3$He cryostats, by a standard four-probe technique.

**High-pressure transport measurement.** For resistance measurement, a non-magnetic BeCu diamond anvil cell (DAC) was used to apply high pressure. A non-magnetic metal gasket was pre-indented to 50 μm thick by the DAC with a pair of 300 μm culet sized anvils. The indented flat area was drilled out and refilled with fine cubic Boron Nitride (cBN) powder. Then the cBN powder was compressed again by DAC and a new hole with a smaller diameter was drilled to serve as an insulated sample chamber. Fine FeS powder, obtained by grinding the single crystals into sub-micrometer size, was filled in the chamber, then a piece of small FeS single crystal was laid down on the top of the prefilled powder. Four Pt leads were used as electrodes to contact single crystal surface without touching each other. All the sample preparation was carried out in a glove box filled with argon gas. The transport measurements were performed in a $^3$He cryostat using the Van der Pauw method. The pressure was determined by the ruby fluorescence method [55].

**High-pressure structure measurement.** To study the structure evolution of pressurized FeS, the *in situ* high-pressure powder angle-dispersive x-ray diffraction (AD-XRD) experiment was performed at room temperature in a Micro x-ray Diffraction beamline (16-BM-D), High-Pressure Collaborative Access Team (HPCAT), Advanced Photon Source, Argonne National Laboratory, using a monochromatic x-ray beam with the incident wavelength of 0.3263 Å. The fine powder sample was pressed into a small piece with the thickness of 10 μm, and filled into the sample chamber shaped by the rhenium gasket. Silicone oil was used as the pressure medium to provide better hydrostatic pressure. The micro-focusing x-ray beam with the full width at half maximum (FWHM) size of 5 μm in diameter was used, and forward diffraction patterns were collected with a two-dimensional mar345 image plate detector. The online Membrane Diaphragm system was used as remote pressure control, together with the online ruby and Microscope system for *in situ* pressure determination. The original XRD patterns were integrated by Fit2D software, and the Rietveld refinement was performed on each XRD pattern to extract the structure evolution information by pressure with the GSAS software.

**Acknowledgements:** This work is supported by the Ministry of Science and Technology of China (National Basic Research Program No. 2012CB821402 and 2015CB921401), NSAF (Grant No. U1530402), and NSFC (Grant No. 51527801), the Natural Science Foundation of China, Program for Professor of Special Appointment (Eastern Scholar) at Shanghai Institutions of Higher Learning, and STCSM of China (No. 15XD1500200). W. Y. acknowledges the financial support from DOE-BES X-ray Scattering Core Program under grant number DE-FG02-99ER45775. We acknowledge Dr. Changyong Park for the assistance of beamline setup during the high-pressure XRD experiments. HPCAT operations are supported by DOE-NNSA under Award No. DE-NA0001974 and DOE-BES under Award No. DE-FG02- 99ER45775, with partial instrumentation funding by NSF. APS was supported by DOE-BES, under Contract No. DE-AC02-06CH11357.

**Author Contributions:** T.P.Y. and Y.J.Y. grew the single crystals. J.Z., F.L.L., Y.X. and M.X.W. performed the high-pressure resistance measurements. J.Z., X.C.H. and L.P.H. analyzed the transport data. F.L.L. and W.G.Y. performed the high-pressure XRD measurements. F.L.L, N.N.L. and W.G.Y. analyzed the XRD data. J.Z., F.L.L., T.P.Y., W.G.Y. and S.Y.L. wrote the manuscript. S.Y.L. and W.G.Y. conceived and designed the project. All authors contributed to the discussion of the results and revision of the manuscript.

**Additional Information:** Correspondence and requests for materials should be addressed to W. G. Yang (yangwg@hpstar.ac.cn) and S. Y. Li (shiyan_li@fudan.edu.cn).

**Competing financial interests:** The authors declare no competing financial interests.

**Figure 1 | Characterizations of FeS single crystals at ambient pressure.**

(**a**) Typical x-ray diffraction pattern of FeS crystals. Only the (00*l*) Bragg peaks show up, indicating that they are well oriented along the *c* axis. The inset shows a photo of the as-grown FeS crystals. (**b**) Single crystal x-ray diffraction spots of a FeS sample. The inset is a scanning electron microscopy image of the surface. (**c**) Temperature dependence of dc magnetization measured in the zero-field-cooled (ZFC) mode at $H$ = 10 Oe parallel to the *c* axis. (**d**) Temperature dependence of the resistivity $\rho(T)$. The inset is an enlarged view of the superconducting transition. The black arrows indicate different definitions of the transition temperature $T_c$. The blue solid line is a fit of the data between 5 and 50 K to the Fermi liquid behavior, $\rho(T) = \rho_0 + AT^2$.

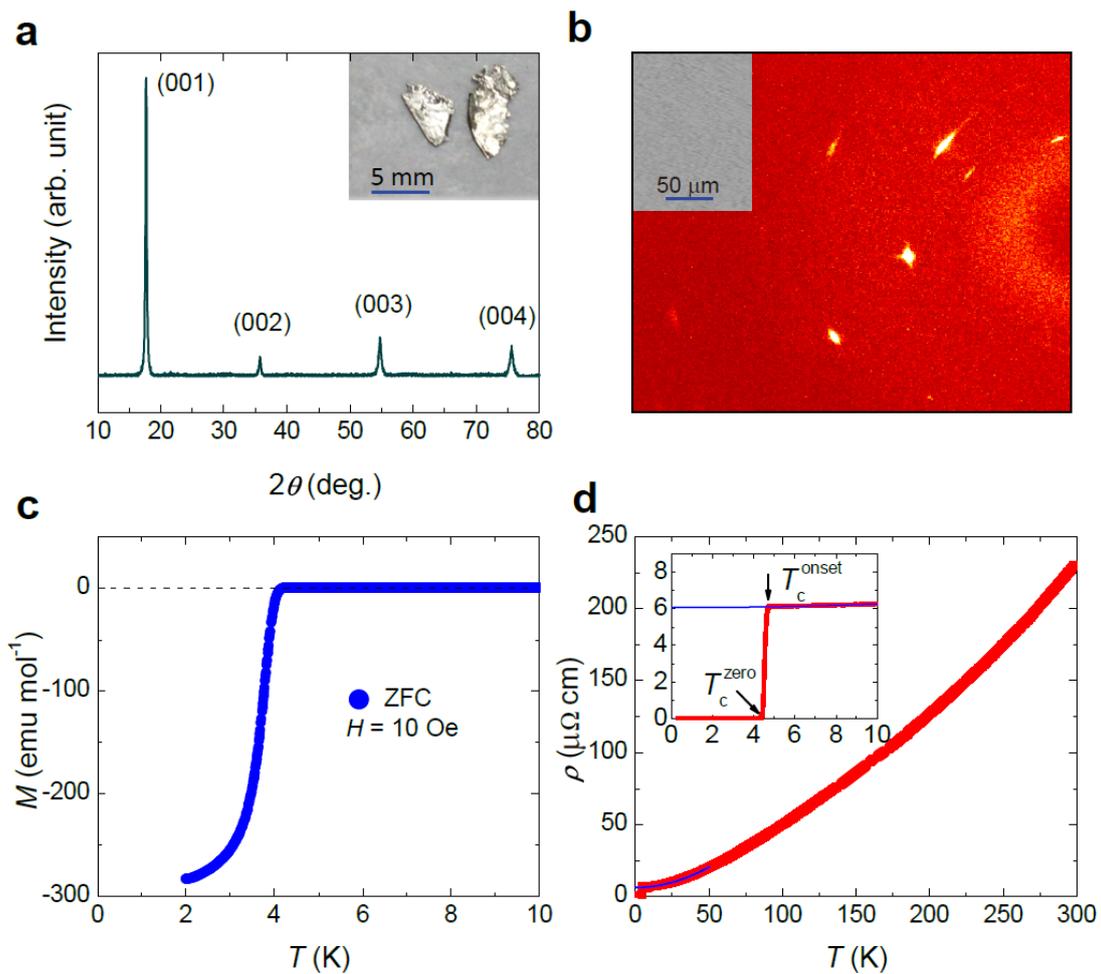

**Figure 2 | Resistance under pressure and magnetic field.**

(**a-c**) The normalized resistance curves of FeS single crystal under various pressures. The black arrows show how the resistance drop evolves with applied pressure. (**d**) The superconducting transition of FeS at 19.0 GPa in various magnetic fields. The inset shows the reduced temperature $T/T_c$ dependence of the upper critical field $H_{c2}(T)$. The violet solid line is a fit to the generalized Ginzburg-Landau model: $H_{c2}(T) = H_{c2}(0)(1-t^2)/(1+t^2)$, where $t = T/T_c$.

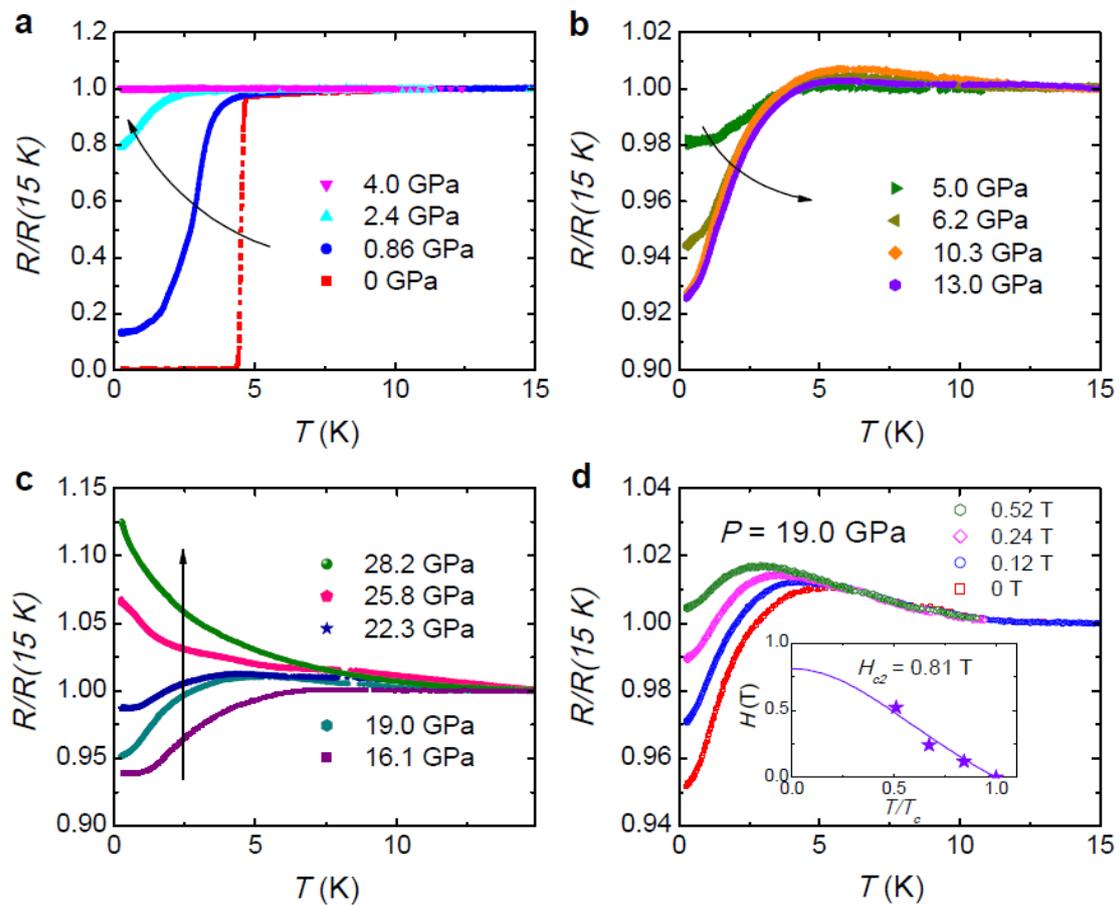

**Figure 3 | Crystal structure evolution under pressure.**

(**a**) The *in situ* powder synchrotron x-ray diffraction patterns of FeS under various pressures at room temperature. The characteristic peaks of two structures are marked to show the evolution with increasing pressure. (**b,c**) Net-like hexagonal structure (*P*-62*c*) and layered tetragonal structure (*P*4/*nmm*) of FeS. (**d-f**) Pressure dependence of the lattice parameters *a*, *c*, as well as unit cell volume. The red and blue solid circles represent the tetragonal and hexagonal phase, respectively. A two-phase coexisting region is highlighted from 7.2 to 9.2 GPa.

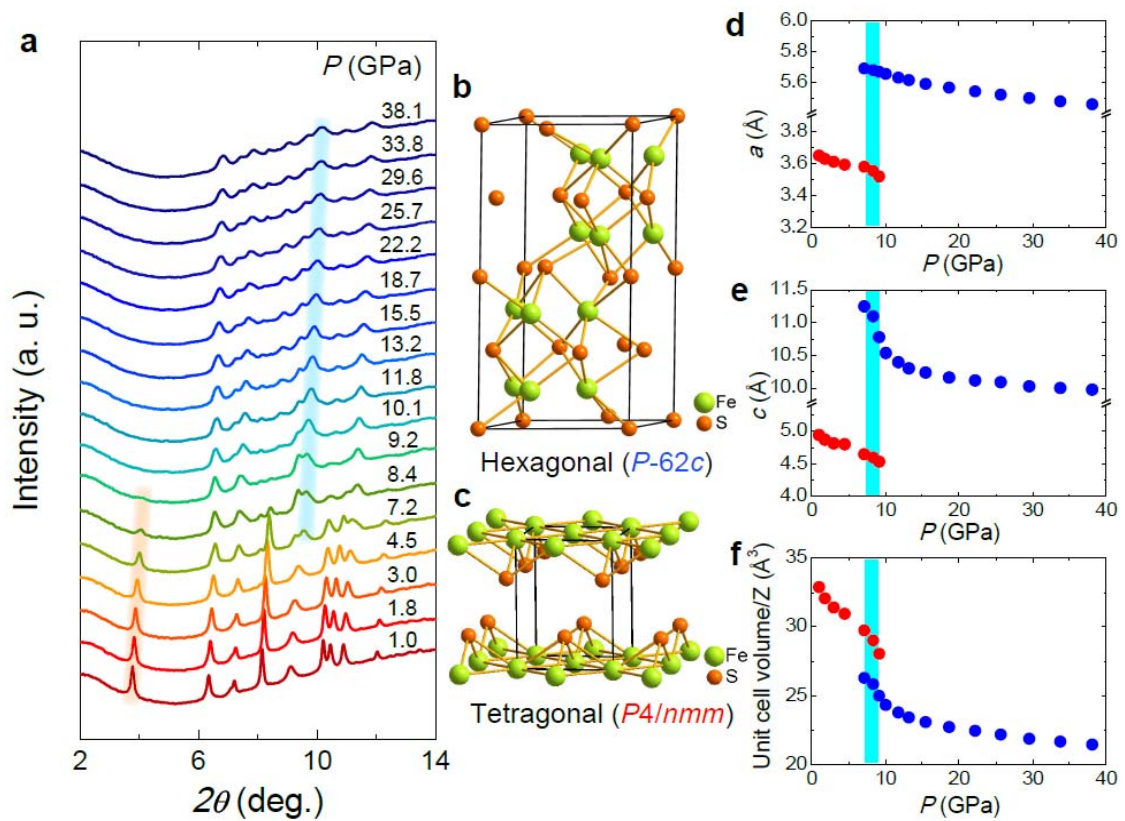

**Figure 4 | Phase contents and temperature-pressure phase diagram.**

(**a**) The pressure dependence of phase contents around the structure transition, which are obtained through refinements. (**b**) Temperature-pressure phase diagram of FeS. There are apparently two superconducting domes, and the second dome is attributed to the remaining tetragonal phase.

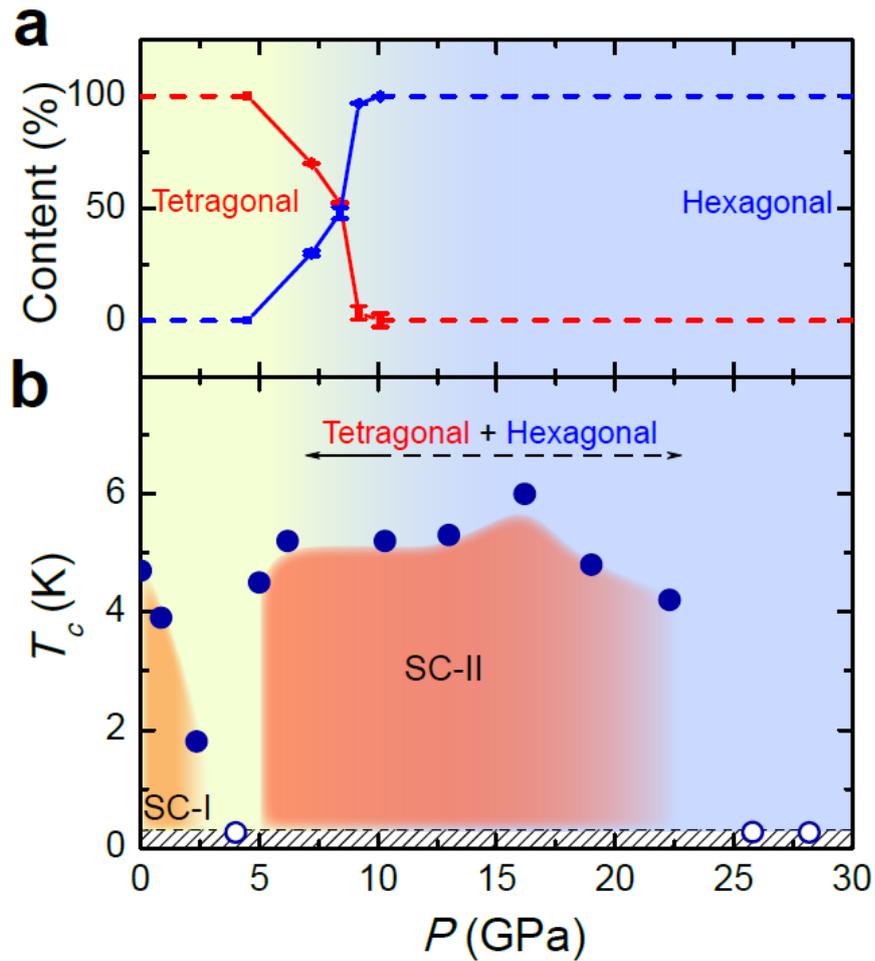

**Figure 5 | Close correlation between structure parameters of tetragonal FeS phase and the superconducting transition temperature under pressure.**

(**a**) S-Fe-S bond angles α, β and Fe-S-Fe bond angles γ and δ. The black arrow marks the anion height. (**b,c**) The pressure dependence of anion height and Fe-S bond length. (**d,e**) The pressure dependence of bond angles α, β, γ and δ, which are obtained through XRD refinements. (**f**) The pressure dependence of $T_c$ from ambient to 10.3 GPa. The similar trend of α angle and $T_c$ indicates a close correlation between S-Fe-S bond angle and superconducting transition temperature.

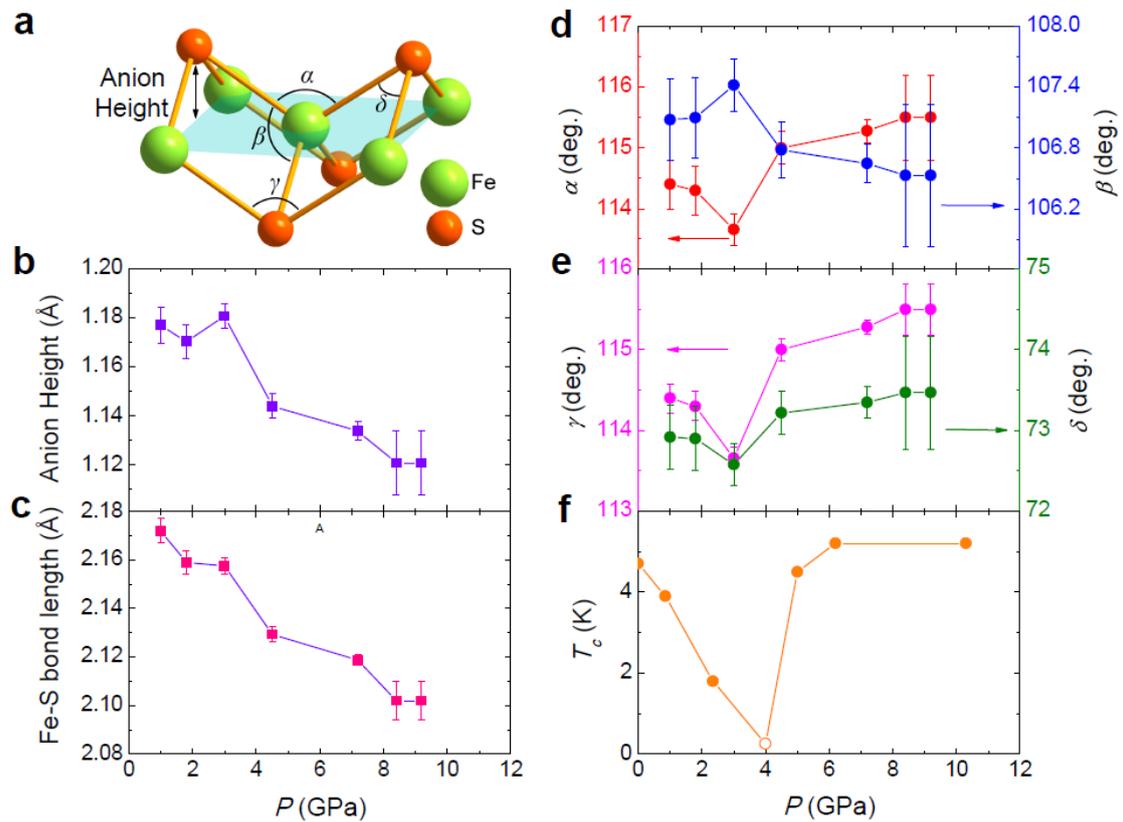